# Impact Of Bike Sharing In New York City

Stanislav Sobolevsky[1*], Ekaterina Levitskaya[1], Henry Chan[2], Marc Postle[2], Constantine Kontokosta[1]

[1]New York University, New York, USA
[2]Future Cities Catapult, London, UK
*Correspondence to be addressed to: sobolevsky@nyu.edu

The Citi Bike deployment changes the landscape of urban mobility in New York City and provides an example of a scalable solution that many other large cities are already adopting around the world. Urban stakeholders who are considering a similar deployment would largely benefit from a quantitative assessment of the impact of bike sharing on urban transportation, as well as associated economic, social and environmental implications. While the Citi Bike usage data is publicly available, the main challenge of such an assessment is to provide an adequate baseline scenario of what would have happened in the city without the Citi Bike system. Existing efforts, including the reports of Citi Bike itself, largely imply arbitrary and often unrealistic assumptions about the alternative transportation mode people would have used otherwise (e.g. by comparing bike trips against driving). The present paper offers a balanced baseline scenario based on a transportation choice model to describe projected customer behavior in the absence of the Citi Bike system. The model also acknowledges the fact that Citi Bike might be used for recreational purposes and, therefore, not all the trips would have been actually performed, if Citi Bike would not be available. The model is trained using open Citi Bike and other urban transportation data and it is applied to assess direct benefits of Citi Bike trips for the end users, as well as for urban stakeholders across different boroughs of New York City and the nearby Jersey City. Besides estimating the travel time and cost savings, the model also reports the associated gas savings, emissions cut and additional exercise for the customers, covering all three areas of anticipated impacts - economic, social and environmental.

## 1. Introduction

Bike sharing systems have gained a lot of popularity in the recent years. New York City's Citi Bike, launched in May 2013, is the largest bike share system in the United States, providing over 605 stations and 10,000 bikes [1] available for the occasional customers and subscribers 24/7. Over 3 years, since the initial deployment until summer 2016, over 25 million trips have been performed. More context on the deployment and its phases could be found in Supplementary Information S1.

Rapidly growing bikesharing systems all over the world have attracted a lot of attention of urban analysts and researchers [2]. NYC DOT Mobility Report states that in New York City's congested Midtown Manhattan, average trips between 1-1.5 miles are more than 5 minutes faster and $10 cheaper by Citi Bike than by taxi [3]. In Montreal, bicycle sharing system in a neighborhood with 12 stations serving 800 meter buffer is expected to increase property value for multifamily housing units by approx 2.7% [4]. In

Minneapolis, a study of Nice Ride system finds correlations between how bikeshare activity increases with the number of food-related businesses within a ⅛ mile walk of a bike share station [5]. Investments in bicycle infrastructure create nearly twice as many direct, indirect and induced jobs per dollar as typical road projects [6].

The motivation of the bikeshare usage has also been studied: 70% of Capital Bikeshare (Washington D.C.) riders choose bikeshare as the quickest and easiest way to get to their destination [7]. Bicycling to work decreases risk of mortality in approximately 40% after multivariate adjustment, including leisure time physical activity [8]. Hubway Bikeshare (Boston, MA) started to pilot programs of subsidized memberships while implementing stations in low-revenue areas in order to increase access and equity of ridership [9]. Cities stand to gain $2.6 billion annually in indirect savings based on lower road construction costs, reduced accidents, and lower carbon dioxide emissions [10].

The reported environmental impacts of bike sharing include reduced carbon emissions and reduction in traffic congestion. Denver B-Cycle users rides accounted in savings of 1,028,836 pounds (466.7 tonnes) of $CO_2$ emissions and 560,424 miles (901,915 km). 43% of B-Cycle rides replaced car trips, resulting in a 15,868 gallon (60,067 liter) decrease in gasoline [11]. Other findings show a reduction in DC traffic congestion of an average 2 to 3% that can be attributed to the presence of a Capital Bikeshare dock (the data includes first three years of system operation) [12].

In New York City several studies have been conducted for the optimization problem of Citi Bike rebalancing system, as well as predicting the demand of bike usage [13-15]. One additional work in New York City focuses on the social impact of Citi Bike in low-income communities [16].

Many studies, especially those focusing on the environmental impact of Citi Bike, simply consider the difference between bike and car ridership, assuming that people would be driving instead. Such an assumption is clearly not realistic, as it could often be preferable, especially in the dense urban environment, to walk or to take public transportation, for example. Not all Citi Bike users even have access to private vehicles. Finally, many Citi Bike trips might be recreational and for that and other reasons not all the trips would have been actually performed, if Citi Bike would not be available. In the present work we construct a balanced baseline transportation choice model to describe projected customer behavior in the absence of Citi Bike with recreational trips taken into account. The open Citi Bike and other urban transportation data are leveraged to train the model and quantify direct benefits of Citi Bike trips in New York City and the nearby Jersey City. The estimated travel time and cost savings are then computed together with the associated gas savings, emissions cut and additional exercise for the customers in order to assess economic, social and environmental impacts.

## 2. The data

Most of the data acquired for the analysis is available publicly through NYC Open Data and websites of public authorities, such as datasets on Citi Bike usage & locations, NYC bike routes (see a map in Figure S2), taxi / Uber usage & locations, subway usage & locations, and NYC-specific data from the U.S. Census Bureau. The detailed description and the links could be found in the Supplementary Information S2.

In particular, Citi Bike publishes large sets of open data on its website, including daily ridership, membership data and trip histories. Usage of taxi is also reported on a detailed level by NYC Taxi & Limousine Commission. Usage of public transit data is limited to the counts of entries and exits at various subway stations over the city. Information on the pedestrian activity and driving personal vehicles is not available at the required scale. In order to fit the parameters of the mobility choice model, the total regular home-work commute demand and preferred transportation options to facilitate it are taken from Longitudinal Employment Household Dynamics and American Community Survey data.

Table 1 shows the usage volume and time per active station in Manhattan, Brooklyn/Queens and Jersey City. Usage is calculated for one full year after each deployment in 2013 and 2015 (only 2015 for Jersey City because bikes were not deployed there in 2013), as well as the comparison year of July 2014 - June 2015, when no additional deployment took place (see a map of deployment phases in Figure S1). The intensity of each station's use heavily depends on the spatial context - the ridership per Brooklyn station is 50% higher in number and 70% longer in total duration than in Jersey City, while ridership associated with an average Manhattan station is 4 times higher than in Brooklyn. The total number of trips does not seem to increase much over time, so on the second year after the deployment it is not higher compared to the ridership over the first year. Based on this, we can focus our impact assessment on one single year immediately after the deployment.

| **Deployment phase** | **Total rides** | **Total time, hours** | **Active stations (on average)** | **Rides per active station** | **Travel time per active station, hours** |
|---|---|---|---|---|---|
| **Jersey City, 2015** | 174,624 | 34,678 | 36.9 | 4621.0 | 931.2 |
| **Brooklyn, 2013** | 536,934 | 128,655 | 78.4 | 6842.6 | 1639.8 |
| **Brooklyn, 2014 (no additional deployment)** | 453,952 | 111,095 | 77.9 | 5834.0 | 1427.8 |
| **Brooklyn/Queens, 2015** | 1,046,191 | 262,542 | 161.2 | 6576.9 | 1654.1 |
| **Manhattan, 2013** | 6,794,155 | 1,443,487 | 249.9 | 27,189.0 | 5776.9 |
| **Manhattan, 2014** | 6,585,370 | 1,376,323 | 249.1 | 26,458.7 | 5530.5 |
| **Manhattan, 2015** | 9,412,124 | 2,152,787 | 285.5 | 33,139.1 | 7573.5 |

**Table 1.** Total volume of Citi Bike usage one year after deployment per deployment area and phase.

Other datasets used in the paper include taxi, Uber and subway usage, as well as the home-work commute data from U.S. Census Longitudinal Employment Household Dynamics. The detailed description of the datasets is included in the Supplementary Information S2. A map with a sample of daily statistics and geographic distribution per transportation mode in New York City can be found in Figure S3.

**3. The single scenario and balanced baseline scenarios for impact assessment**

Each Citi Bike station on average facilitates thousands and sometimes even tens of thousands of trips per year. The first direct impact of CitiBike to consider is the savings in transportation time and expenses for the customers who otherwise would have to use alternative (often slower or more expensive) transportation modes, unless the trip could have been skipped in case a bike is not available.

In order to estimate the impact above, we need to understand how/if customers would have performed the same trips without having bike share as an option.

While not realistic, a single scenario analysis replacing Citi Bike with one single alternative, such as driving, could be helpful to understand the upper bounds for the impacts. Assuming that everyone would drive instead of taking a Citi Bike in Manhattan during one year from July, 2013 until June, 2014, given the cumulative bike travel distance to be replaced by driving, one could come up with an approximate estimate of 4,996 liters of gas per year saved by an average station deployed (by considering the average optimal route driving mileage provided by Google Maps API for the reported actual bike trips (see Supplementary Information S3) and related average car gas consumption). Assuming that everyone would walk instead gives 3,204 hours saved per station annually (based on travel distance, average walking speed with respect to averaged street topology and actual bike travel time recorded). Assuming that everyone would take a taxi gives $363,411 of annual savings per station (based on travel distance, time and its cost, as well as taxi fares). Results for all single scenarios are reported in the Table 2 below. Calculations assume normal taxi fares and wait time, average private vehicle car gas consumption and maintenance costs, average vehicle, pedestrian and public transit speed as per tables S1-S3 from appendix S3. Economic effect of time savings is incorporated into cost savings with respect to average wage.

| **Scenario** | **Travel cost savings, $** | **Gas savings, liters** | **Travel time savings, hours** |
|---|---|---|---|
| **All driving** | -32,237 | 4,996 | -897.3 |
| **All take taxi** | 363,411 | 4,996 | 2274.8 |
| **All walking** | 191,199 | - | 3,204 |
| **All public transit** | 24,012 | - | -191.9 |

**Table 2.** Impact assessment for the scenarios of replacing all bike trips with a single selected transportation mode.

Each scenario (assuming that all Citi Bike trips have replaced a given transportation mode) allows us to focus on specific gains related to such a replacement and provide a spectrum of possible gains(together with the extreme estimates). In reality, different bike trips have replaced different alternative modes, so the resulting impact is composite, providing a certain balance of the benefits above. This is due to different optimal transportation choices for different types of trips, as well as limited availability of certain transportation, like private vehicles (average car ownership among the households in most areas of NYC and Jersey City varies between 20 and 60%). Below we construct a more balanced baseline scenario, estimating a combination of alternative transportation modes which could have served as a best choice given the actual bike trips.

Based on it, the following scheme for the impact assessment approach is proposed on the Figure 1 below.

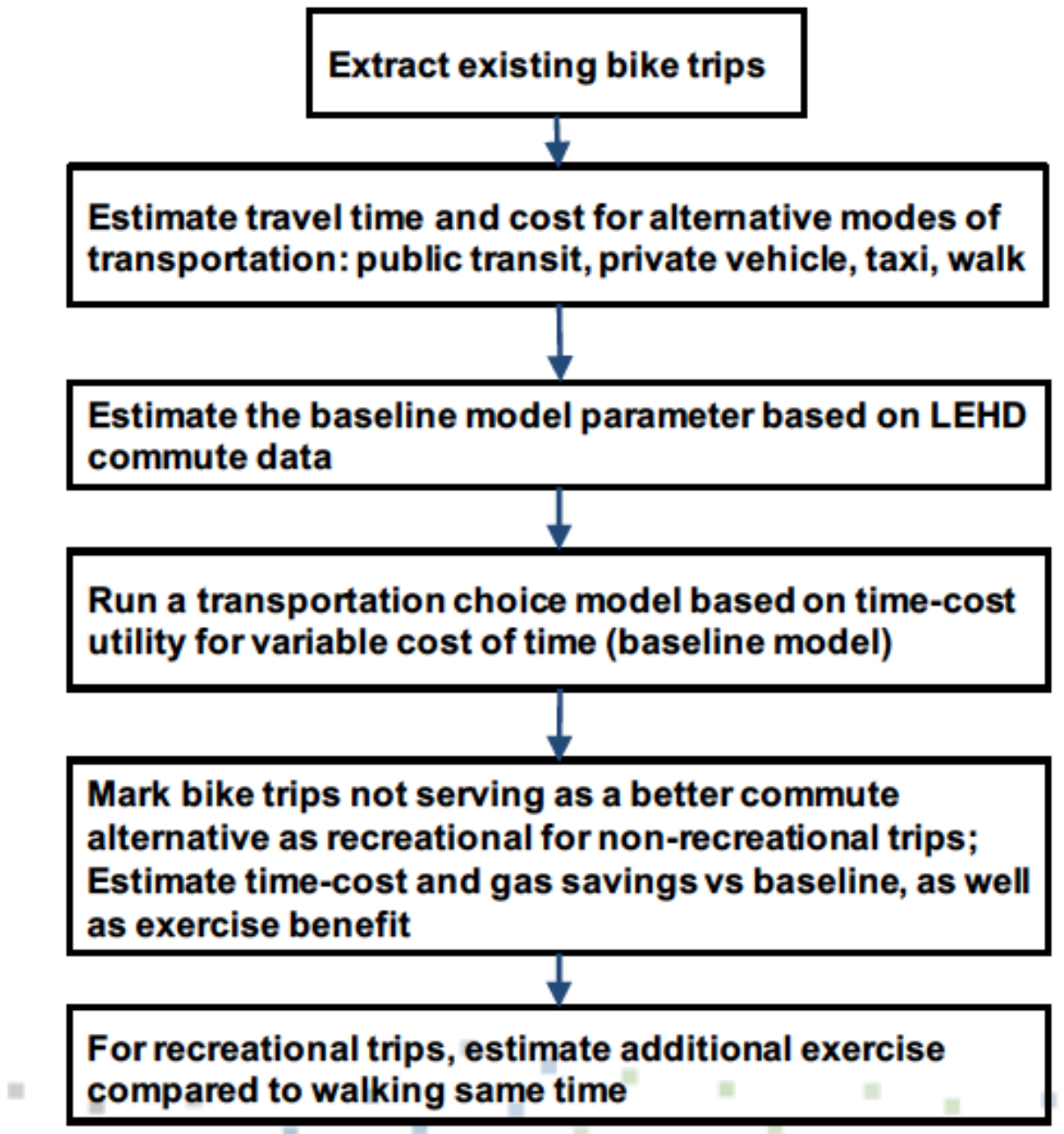

**Figure 1.** Conceptual scheme of the impact assessment.

The impact assessment is performed based on all the Citi Bike trips across NYC and Jersey City (a nearby town in New Jersey across the Hudson river) as provided by the data, which includes time, beginning and the end of the trip.

**4. The baseline model**

4.1. The general modelling framework. The baseline model considers main available transportation alternatives for each Citi Bike trip - walking, public transportation, private vehicle (for those having access to them) or for hire vehicle (FHV) - and estimates the probabilities of taking each (Figure 2).

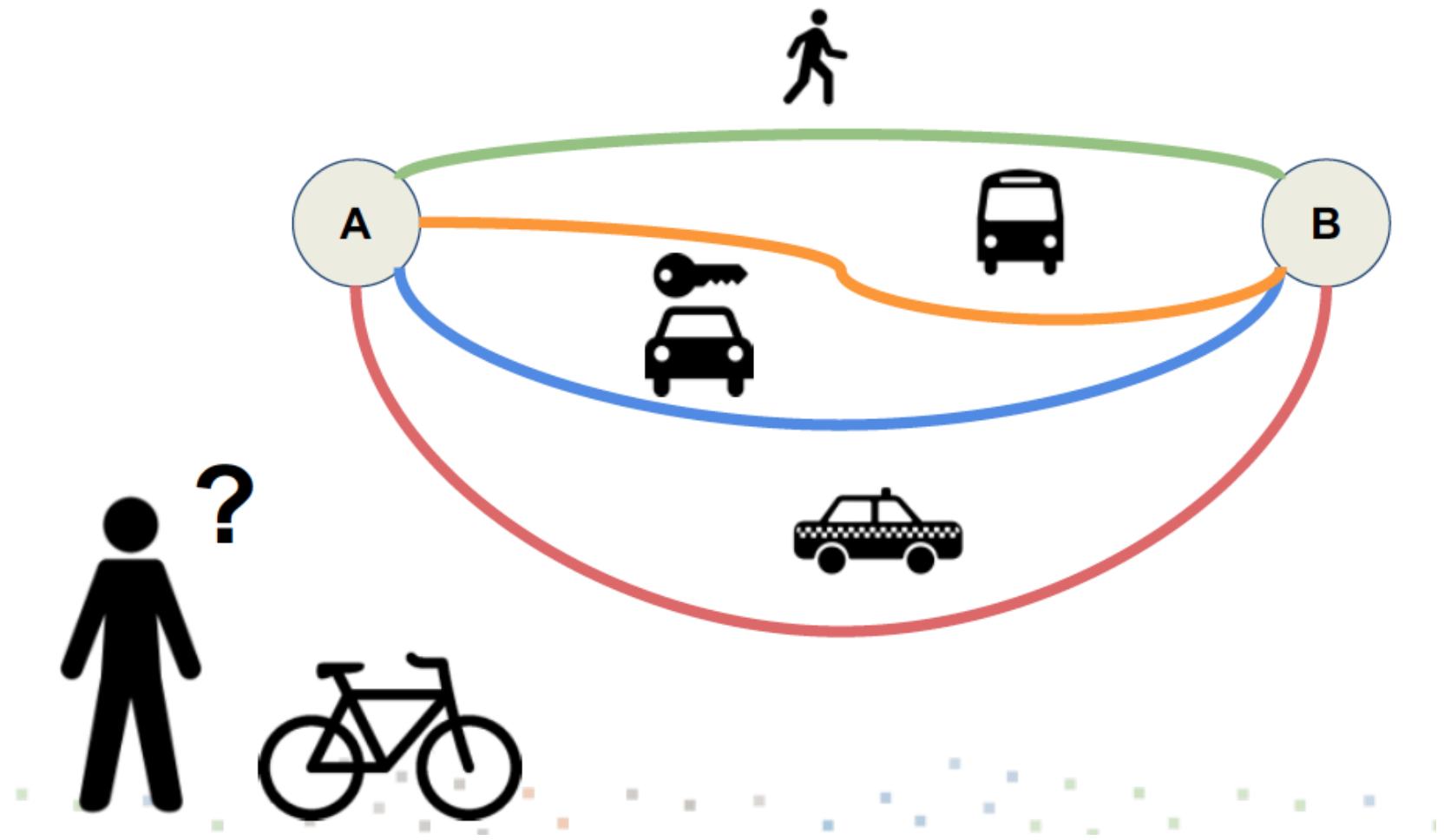

**Figure 2.** Choosing a transportation mode to replace a Citi Bike trip.

A classic approach in the modeling of transportation mode choice is a multinomial logit (MNL) discrete choice model [17-19] assigning a certain utility score or cost (a negative utility) $U_j(t)$ to each transportation option $j$ for a given trip $t$ and to define a probability $P_j$ of choosing an option $j$ as

$$P_j(t) = \frac{e^{-\lambda U_j(t)}}{\sum_k e^{-\lambda U_k(t)}} \quad (1)$$

where $\lambda$ is the utility weight coefficient. While different transportation modes provide different sets of costs and benefits to the user, the optimality of each could depend on the individual trade-off between those costs and benefits. A major decision to be made is whether to prefer faster but more expensive transportation or slower but more affordable. The simplest way of defining the utility of the transportation mode is by converting both time and cost to monetary values, e.g. through valuing time at the level of an average hourly wage for the commuters. This way, the utility score of the trip combines its direct cost to the customer $C_j(t)$ and the individual perception of the cost of time $T_j(t)$ spent on a trip as

$$U_j(t) = C_j(t) + wT_j(t) \quad (2)$$

where $w$ is the random variable providing an individual cost of a unit of time (estimated based on the income distribution in the considered area). Because of the uncertainty of the quantities $T$ and $C$ included in the above assessment (the actual travel time and cost might vary depending on the actual traffic conditions), random variables might be considered rather than constant values.

<u>4.2. The choice of the coefficient $\lambda$.</u> The model is sensitive to the choice of the coefficient $\lambda$. We choose it by fitting the model to the known overall morning commute around the city, according to Longitudinal Employment-Household Dynamics (LEHD) data which provides origins and destinations of regular home-work commutes, and evaluate the results against American Community Survey (ACS) data which provides some ground truth for the preferred transportation mode of people's commute. Specifically, for each pair of origin and destination census blocks the number of home-work commuters is provided and for each borough (county) people report their preferred mode of transportation (driving, FHV (taxi), public transportation, walking). This is a different portion of human mobility from the Citi Bike ridership that the impact assessment is focusing on. LEHD home-work commute is used exclusively for fitting the coefficient $\lambda$.

For each origin-destination pair and transportation mode we estimate utilities (2) based on the distance between them, as well as the expected length of the route and travel time for each transportation mode facilitating such a distance. Travel time is estimated based on the average travel time/distance per mile of geographical distance according to Google Maps API, as reported in Supplementary Information S3 for different times of the day/week. Driving and taxi travel time are estimated using a combination of the normal and pessimistic scenarios. Then, the transportation time-cost for each transportation mode is assessed in the following way: 1) for walking, it is simply the cost of time according to the anticipated walking time between the origin and destination of the trip, 2) for public transportation, it is the fare multiplied by the average fraction of pay-per-ride customers (taken as 48% here, as 52% of the customers use passes [20], plus cost of transportation time, plus cost of an average wait time), 3) for taxi, it is the initial charge, plus mileage/time fare, plus tips, plus cost of time for the passenger divided by the average number of passengers per ride (according to NYC TLC taxi data (Supplementary Information S2), around 1.7), 4) for private vehicle driving (available for the percentage of residents having cars in their possession), it is an American Automobile Association estimate [21] for an average per mile cost of gas, plus maintenance, plus depreciation, which is around 60 cents per mile, divided by the average number of passengers per ride, plus the parking cost.

The above estimate depends on some assumptions which need to be made in view of the limited amount of ground truth data, such as the balance between standard and pessimistic driving times (varying between 0 for all standard to 1 for all pessimistic) and the effective cost of parking (varying from 0 to 8 hours of the average parking rate, as, while people driving to work need to park their car for the day, many have free to compensated parking from their employers). Varying those parameters $x$ results in additional uncertainty in the estimate for $\lambda$ which is described below.

For each choice of $\lambda$ and the parameters $x$, the model (1) provides transportation mode distribution to be matched with the actual data from ACS for the entire NYC. Specifically, it gives a probability distribution for the total number of the trips $\mathrm{T}_j(\lambda, x)$ (with a standard deviation $\sigma_j(\lambda, x)$; the distribution is close to normal) to be facilitated by the transportation mode $j$ and allows to assess the likelihood of the observed ridership volume $T_j^*$ as

$$L(\mathrm{T}^*|\lambda) \sim \int e^{-\sum_j (T_j^* - \mathrm{T}_j(\lambda,x))^2/\sigma_j(\lambda,x)^2} dx$$

The Figure 3 below shows the goodness of fit metrics (in terms of Dice [22] and Cosine [23] similarity) for each $\lambda$ averaged over all the values of parameters $x$, obtained through simulation of various $\lambda$ and $x$ and comparing the predicted distribution of transportation modes with the reported one.

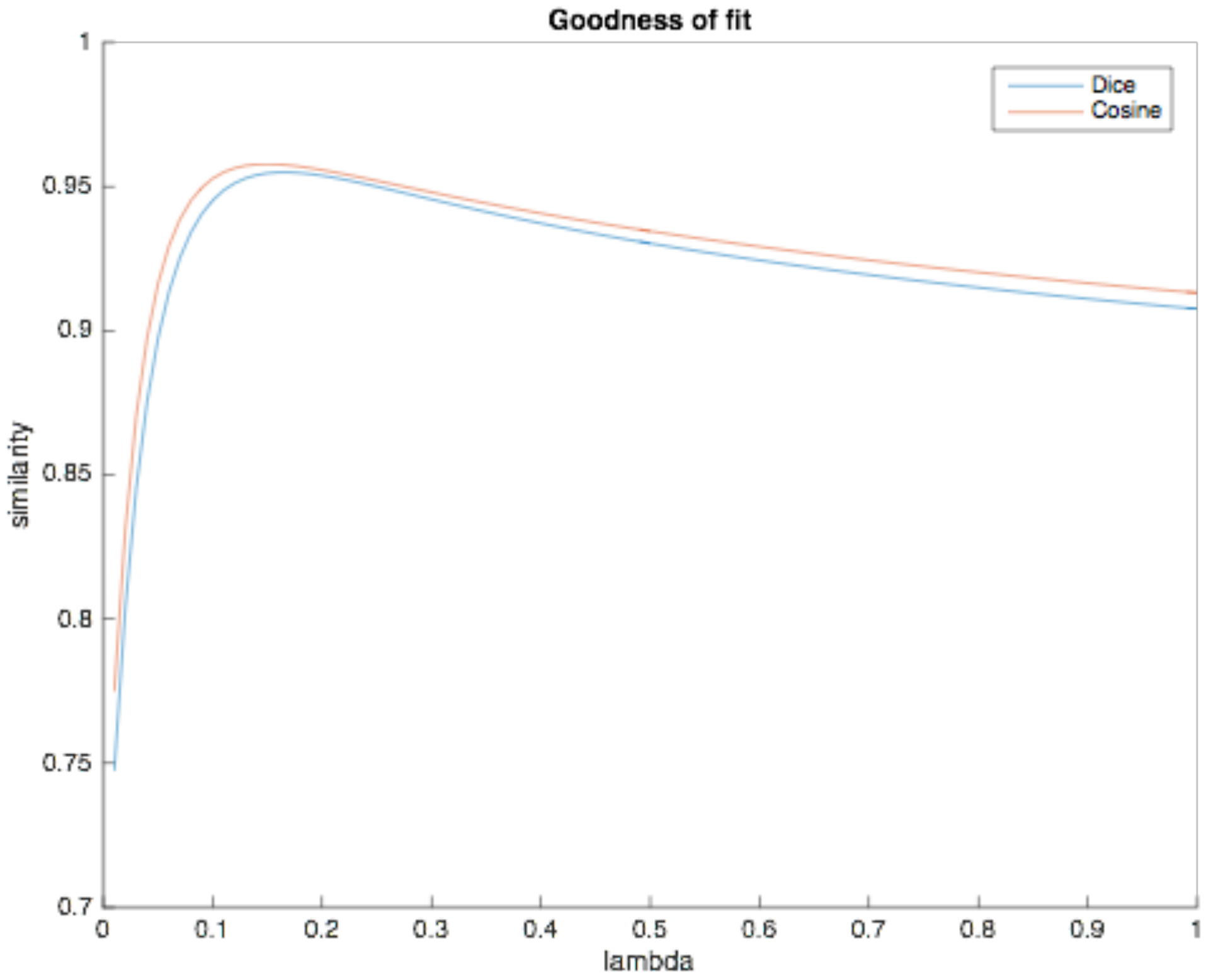

**Figure 3.** Goodness of fit for each $\lambda$ in terms of Dice and Cosine similarity between the predicted distribution of transportation modes and the reported one.

Then, using a Bayesian inference framework, one can find

$$p(\mathrm{T}^*|\lambda) \sim L(\mathrm{T}^*|\lambda)p(\lambda).$$

Going with an uninformed prior $p(\lambda) \sim const$ and considering $\lambda \in [0,1]$ (as we see below, higher values of $\lambda$ have neglectably low likelihood), we get a probability distribution for $\lambda$, as shown in the Figure 4 below.

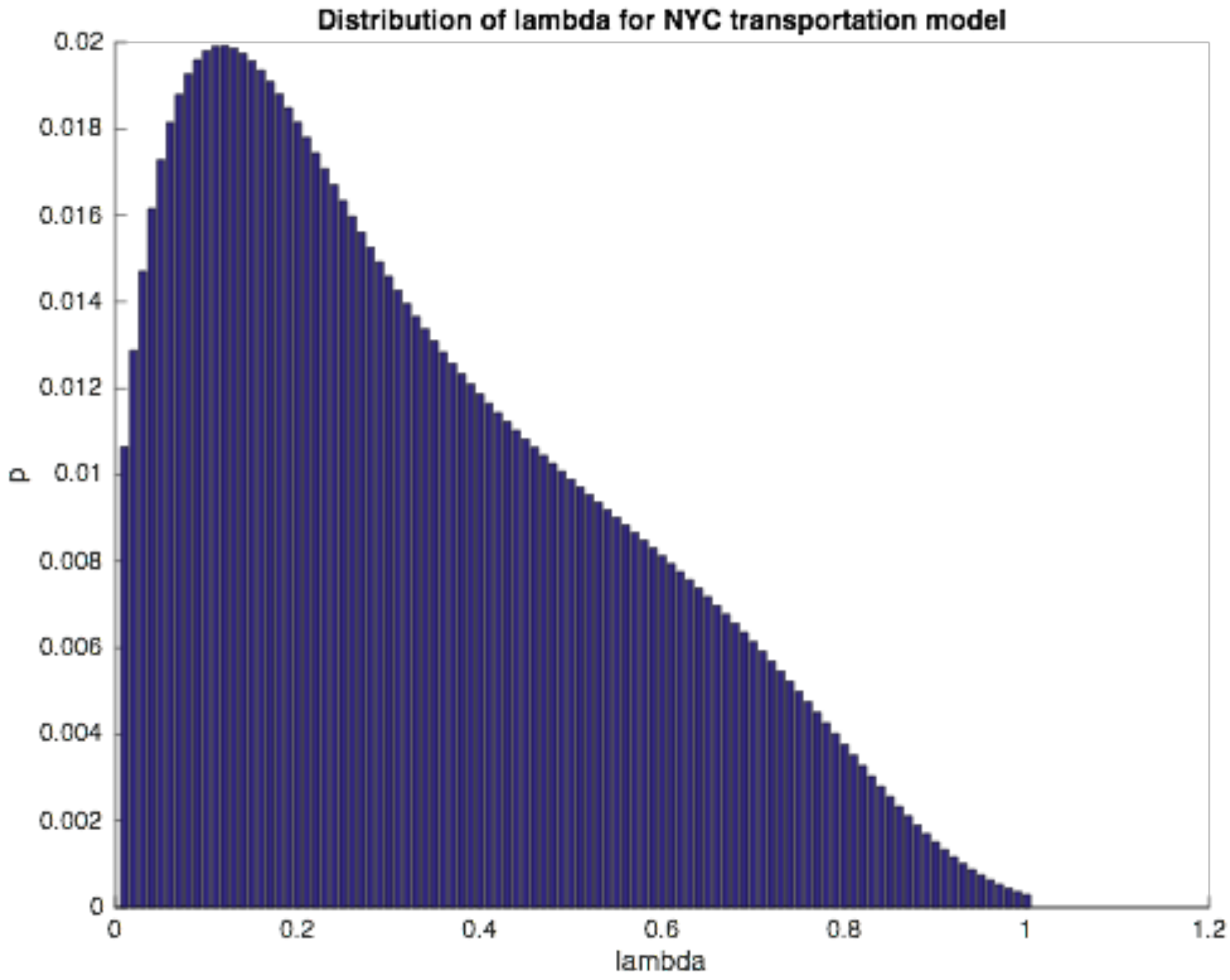

**Figure 4.** Distribution of $\lambda$ for NYC transportation model.

The max-likelihood estimate is $\lambda = 0.12$ for NYC, being close to the value $\lambda = 0.128$ of the travel cost weight reported in [24]. Below we start from using this best estimate of $\lambda = 0.12$, but also assess uncertainty of the resulting estimates by simulating different values of $\lambda$ , according to the above probabilistic distribution.

Now, once the parameter of the model (1) is estimated, we apply it to estimating the baseline scenario: which transportation modes would the people have used if Citi Bike were not available, enabling impact assessment with respect to additionality.

4.3. Estimating the baseline scenario. For each observed bike ride we consider available alternatives - walking, taking public transit, a FHV (taxi) or driving a private vehicle (conditional to having access to a private vehicle, introduced as a binary random variable with a Bernoulli distribution having its mean set at the area's car ownership rate) and estimate their time-cost utility $U_j$ (where $j$ goes through all the available options) as described above (however, for the driving we do not include day parking, as it was done for home-work commute). Utilities $U_j$ are random variables incorporating uncertainty of the travel time estimate. Then, the model (1) provides probabilities $P_j(t)$ of using each alternative transportation mode to replace each considered bike trip, if Citi Bike is not available.

The utility $U_{bike}$ of using a Citi Bike for the trip is defined as the cost of time according to the data record (subscription price is not included into the analysis, as this is a fixed annual payment which, once made, does not affect the transportation choices - after that, the rides are perceived at zero marginal cost; it is included as a fixed cost for the customers to be evaluated versus their projected benefits). The utility of taking Citi Bike is compared against each of the alternatives $U_j$, giving comparative benefits of $B_j(t) = U_j(t) - U_{bike}(t)$.

The marginal benefit of a Citi Bike trip *t* replacing other transportation means is then defined as a random variable taking each of the values $B_j(t)$ (being also random variables themselves) with a corresponding probability $P_j(t)$. The average estimate for the benefit of the trip $E[B(t)]$ is then defined as

$$E[B(t)] = \sum_j P_j(t) E[B_j(t) | B_j(t) > 0] = \sum_j \frac{e^{-\lambda U_j(t)}}{\sum_k e^{-\lambda U_k(t)}} E\left[U_j(t) - U_{bike}(t) \middle| U_j(t) > U_{bike}(t)\right].$$

In cases when $U_j(t) \le U_{bike}(t)$, i.e. the bike ride does not appear to be a better commute choice compared to the considered transportation alternative $j$, we consider this ride to be a recreational trip with a corresponding probability $P_j(t)$ (although the intended purpose of the Citi Bike is to provide commute options, many people still use it for recreational purposes which might take longer than commute trips for the same origin and destination; we assume recreational purpose whenever the trip does not seem to be a rational commute choice).

When we assume a recreational trip rather than commute, economic benefit for the passenger is irrelevant. For recreational trips, we assume that if bike is not available people would take a recreational walk for the same amount of time.

Our assumption that all the rational choice trips should be replaced with other transportation modes if Citi Bike were not available should be acknowledged as one of the limitations of the model. In reality, the availability of Citi Bike as a commute option might enable certain trips, which otherwise would not have been conducted at all. Strictly speaking, the marginal benefits of such trips should not be counted towards the impact of Citi Bike, however, the benefit of being able to make those trips could be added instead.

The Table 3 and Figure 5 below report the outcomes of the transportation choice modeling providing the ways Citi Bike trips are likely to be replaced if bikeshare was  not available.

| **Deployment phase** | Total Citi Bike ridership | Presumably recreational | Replaced by walking | Replaced by public transit | Replaced by taxi | Replaced by private vehicle |
|---|---|---|---|---|---|---|
| **Jersey City, 2015** | 174,624 | 61,224 | 37,094 | 44,372 | 25,857 | 6,076 |
| **Brooklyn, 2013** | 536,934 | 155,532 | 108,977 | 161,010 | 66,190 | 45,225 |
| **Brooklyn/Queens, 2015** | 1,046,191 | 282,005 | 235,734 | 305,433 | 139,872 | 83,148 |
| **Manhattan, 2013** | 6,794,155 | 1,986,272 | 1,434,988 | 2,364,693 | 754,566 | 253,636 |
| **Manhattan, 2015** | 9,412,124 | 2,489,291 | 2,081,333 | 3,403,722 | 1,054,500 | 383,278 |

**Table 3.** Projected replacement of Citi Bike ridership with alternative transportation modes in different deployment areas/periods (average estimates for $\lambda = 0.12$).

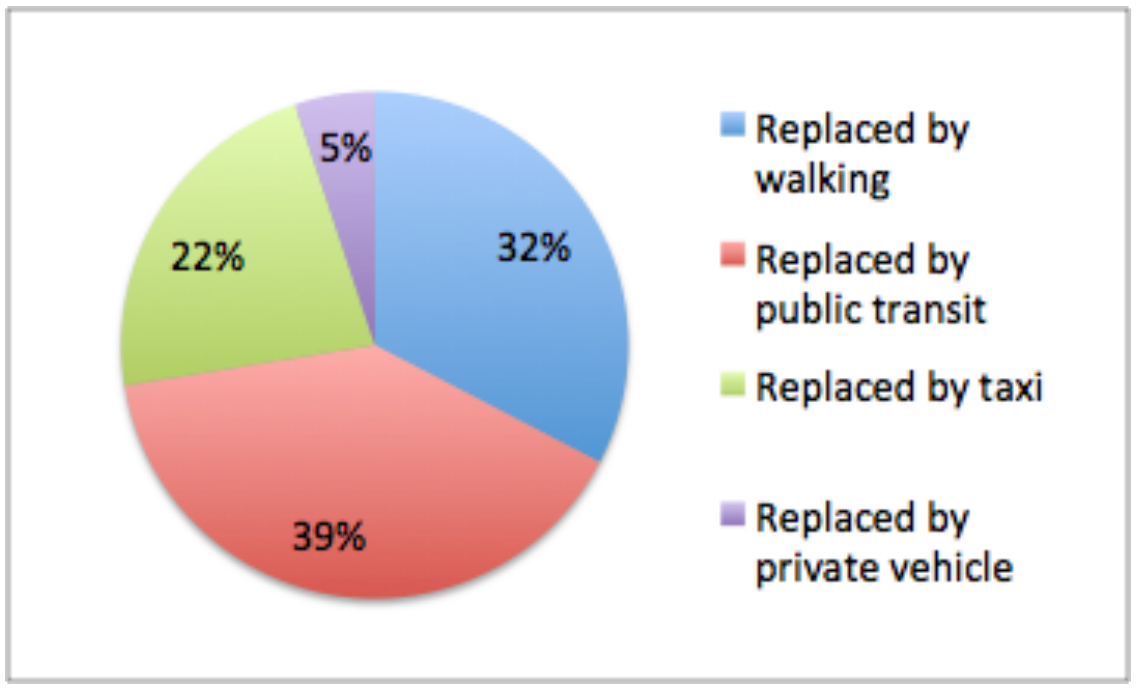

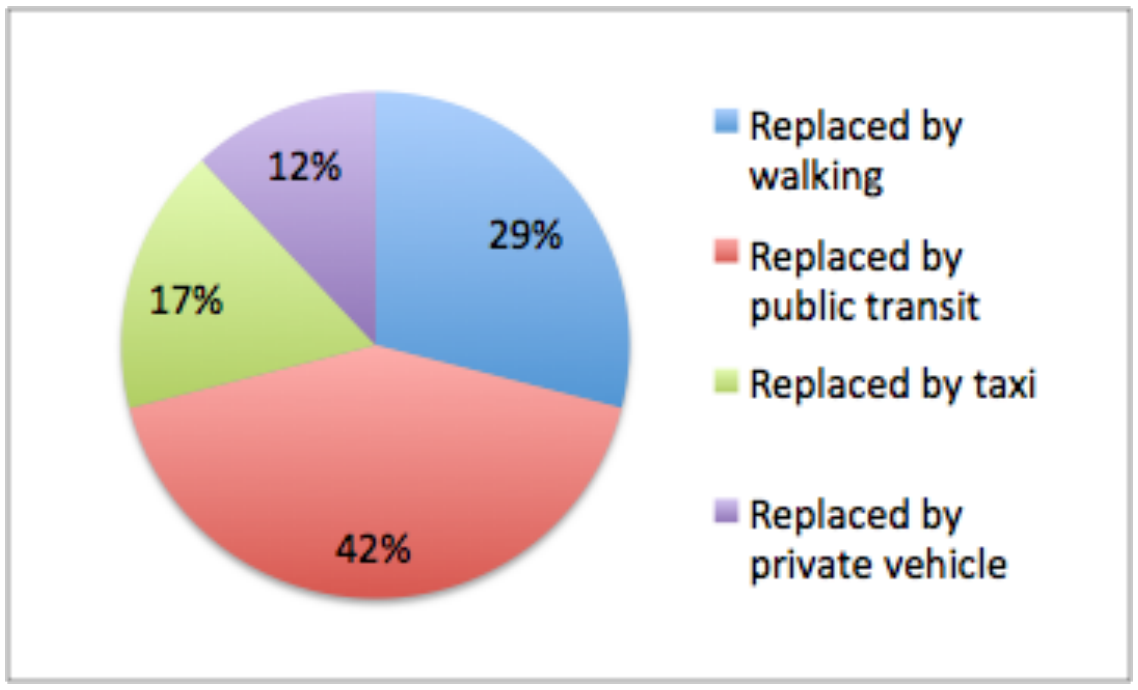

**Jersey City, 2015** **Brooklyn, 2013**

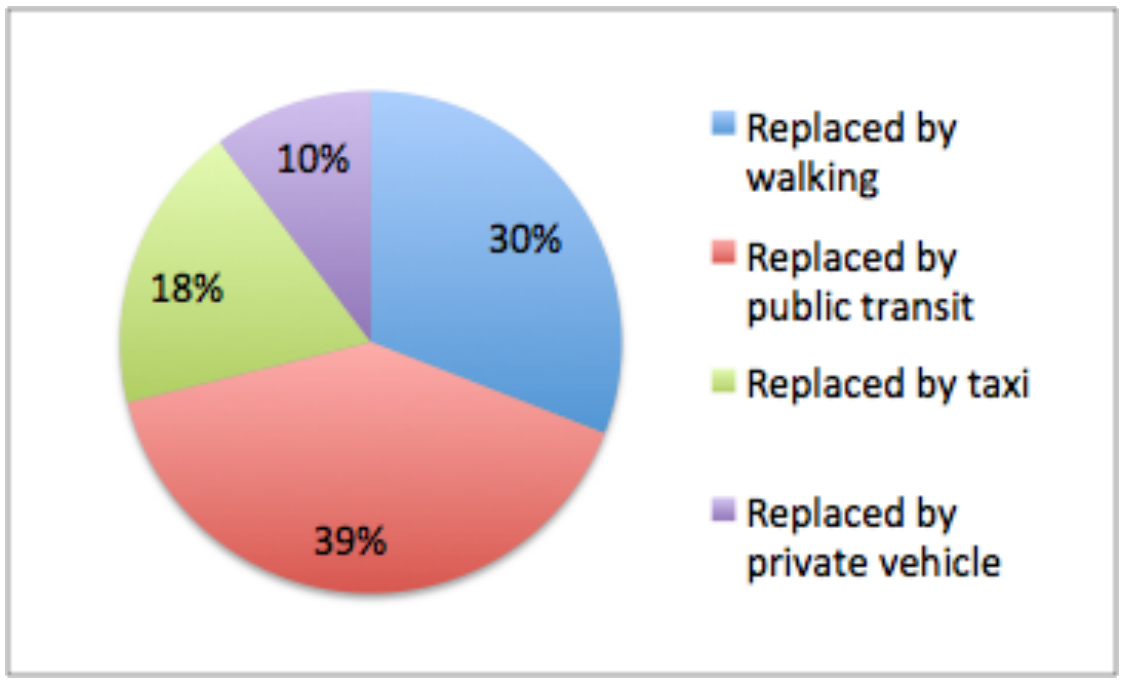

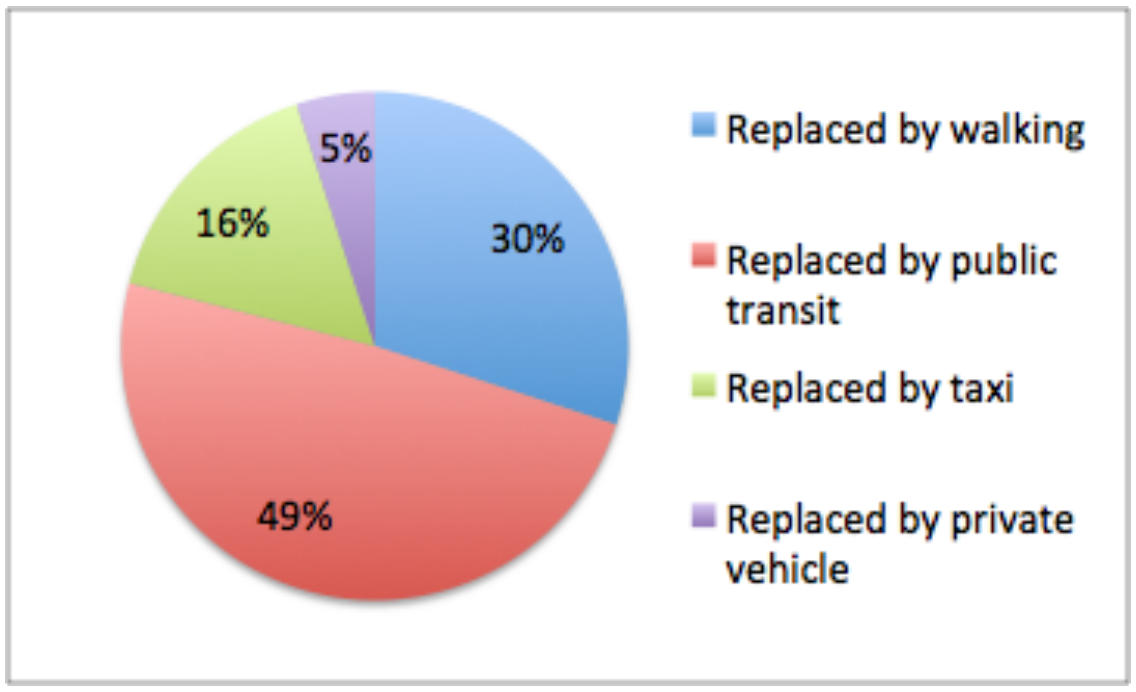

**Brooklyn/Queens, 2015** **Manhattan, 2013**

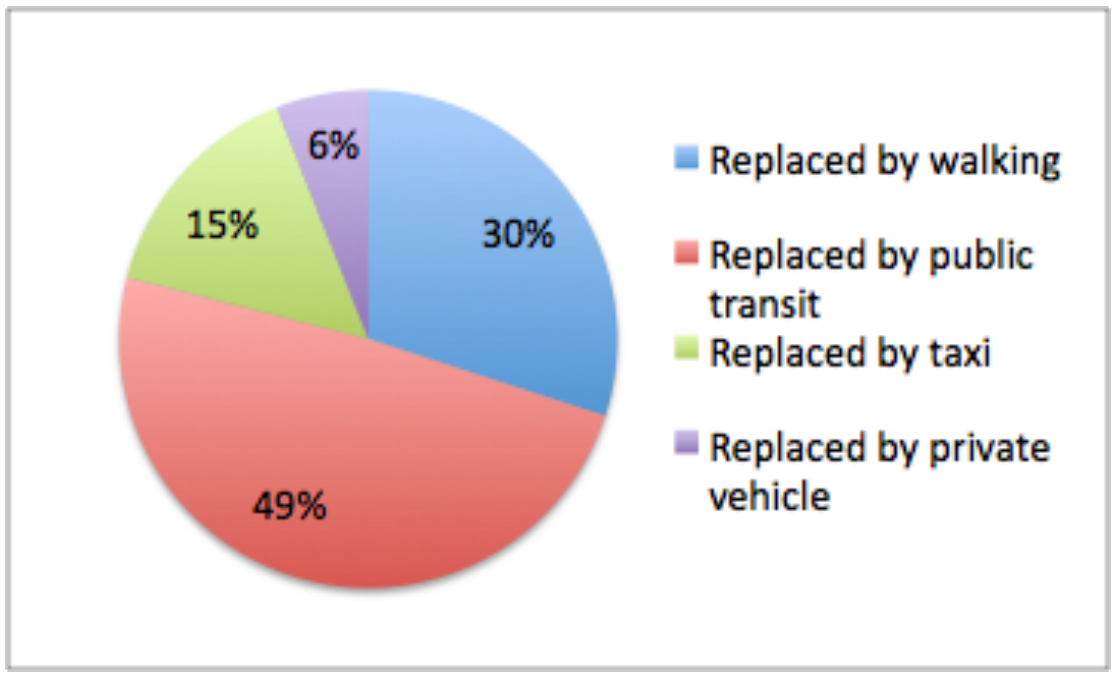

**Manhattan, 2015**

**Figure 5.** Graphical representation of the projected replacement of Citi Bike ridership with alternative transportation modes in different deployment areas/periods (average estimates for $\lambda = 0.12$).

**5. Economic benefit**

The model above allows to estimate the direct economic marginal benefit for the Citi Bike riders in terms of travel cost+time savings versus alternative transportation modes, while separating recreational trips where no economic effect is relevant from commute trips where such cost+time savings are likely to arise.

| Deployment phase | Economic benefit, commute $ | Health benefit, exercise calories | Travel time savings, hours | Environmental benefit, gas saved, liters |
|---|---|---|---|---|
| **Jersey City, 2015** | **11513.82**<br>8965.30±2813.01: [3451.80..14478.79] | **364066.1**<br>356553.9±14796.7: [327552.3..385555.5] | **210.3**<br>171.7±32.2: [108.5..234.9] | **86.2**<br>90.1±15.6: [59.5..120.7] |
| **Brooklyn, 2013** | **38180.95**<br>32976.29±6441.95: [20350.06..45602.52] | **800523.4**<br>775622.7±25114.2: [726398.8..824846.6] | **1075.7**<br>989.1±80.9: [830.4..1147.7] | **478.6**<br>511.8±38.0: [437.3..586.3] |
| **Brooklyn/Queens, 2015** | **26104.17**<br>21415.49±5345.74: [10937.84..31893.14] | **765498.4**<br>742969.9±22692.2: [698493.2..787446.6] | **642.8**<br>579.7±58.7: [464.7..694.7] | **330.2**<br>335.2±32.3: [272.0..398.4] |
| **Manhattan, 2013** | **119475.04**<br>97833.70±29893.81: [39241.84..156425.56] | **2789504.5**<br>2712856.9±88528.7: [2539340..2886373] | **1887.1**<br>1611.1±326.8: [970.7..2251.6] | **675.4**<br>536.1±277.7: [-8.2..1080.4] |
| **Manhattan, 2015** | **146514.83**<br>119528.88±36865.08: [47273.32..191784.45] | **3609389.9**<br>3524018.5±98688.5: [3330589..3717448] | **2301.1**<br>1954.0±407.7: [1154.9..2753.2] | **839.6**<br>672.8±334.2: [17.8..1327.9] |

**Table 4.** Annual economic, social (health and travel time) benefits for commuters as well as environmental benefit (gas savings) per station for each deployment phase. For each quantity the max-likelihood as well as the unbiased average estimate together with standard deviation and the confidence interval are reported.

First column of the Table 4 provides the values of the estimated economic benefit together with the uncertainty of the estimate (standard deviation and 95% confidence interval) due to the probabilistic nature of the model. Economic impact is dramatically different for different areas - even more than the ridership amount is. Manhattan deployments show substantial annual economic impact of over $100.000 per each station deployed, while the stations in Brooklyn and Jersey City show more modest impact of ten to a few tens of thousands each. This can be attributed to the fact that the bike sharing deployments largely depend on the network effect [25, 26] - new stations bring more benefits in the areas with a denser population and a denser existing Citi Bike network. Comparing the impact of initial deployment phase in 2013 with the subsequent deployments in 2015 shows that economic benefit reaches its full capacity in Brooklyn during the first year after the initial deployment. Two years later, even after additional deployments further strengthening the network, the effect of an average Brooklyn station is even lower than the initial impact in 2013. Impact per station in Manhattan keeps growing and in 2015 is 20% higher than in 2013, which could be attributed to the growing network coverage providing additional ridership options. In both cases, the first post-deployment year could be already used to get a rough estimate for the anticipated long-term effect.

## 6. Other benefits

While economic benefit incorporates both savings in direct costs and travel time, we think that focusing on the travel time savings specifically provides an additional social perspective, as this time might be made free for social activities. Projected total travel time savings for the users are reported in the third column of the Table 4 and vary from 185 hours per year saved per average station in Jersey City to around 1000 in Brooklyn and around 2000 hours saved for each station in Manhattan (which seems consistent with the general understanding of the density of those areas). It is important to note again that this analysis excludes recreational trips, where travel time saving considerations are irrelevant.

The usage of the Citi Bike also causes an additional health benefit to the riders. Depending on the best alternative transportation mode, the additional exercise benefit can be estimated at an average of 650 calories per hour for an average rider [27], if walking is not the best alternative and taking into account the difference between ridership and walking exercise (411 calories per hour otherwise [28]). As walking usually takes longer, the amount of calories from biking might not necessarily exceed the amount of calories from walking along the same route. For presumably recreational trips we consider the exercise benefit based on the assumption that the same amount of time would be spent on a recreational walk otherwise. Overall, the amount of exercise coming from replacing other transportation modes and/or recreational trips is significant. It is important to note again that the exercise benefit from Jersey City deployments differs from the top benefit at Manhattan by an order of magnitude, like all other benefits.

Replacing the usage of private cars or taxies creates additional environmental benefit in gas savings which causes corresponding cut in the vehicular emissions. Each time a bike trip is considered to replace a private vehicle or taxi ride, the gas benefit is estimated with respect to an average gas consumption of 25 miles per gallon [29] (9.5 l/100 km). According to the model estimate, each station saves on average a moderate amount of gas between 85 liters a year in Jersey City to 840 liters a year in Manhattan. Those savings can be directly translated into saved vehicular emissions of 0.2 to 2.0 metric tons of carbon dioxide per year accordingly [30].

Results for other types of emissions in accordance with [31] are reported in Table 5.

| Engine Type | EF Units | **PM** [a] | **$PM_{10}$** [a] | **$SO_2$** [b] | **$NO_X$** [c] | **VOC** | **CO** [c] | **CO2** |
|---|---|---|---|---|---|---|---|---|
| Gasoline | lb/mm btu [d] | 0.1 | 0.1 | 0.08 | 1.63 | 2.1 | 0.99 | 154 |
| | | **0.436** | **0.436** | **0.348** | **7.110** | **9.161** | **4.31** | **671.822** |

**Table 5**. Estimation of different types of emissions from gas savings (metric tons).

## 7. Discussion

By adopting a bike sharing scheme the New York City Department of Transportation (NYC DOT) was looking to reduce emissions, road wear, collisions, and road and transit congestion and looking to improve public health (this is the project's theory of change) [32]. With available data, we are able to estimate the following aspects:

- Gas saved translated into saved emissions (environmental)
- Calories burnt by individuals, which improves public health (social)
- Increase in commercial activity for local businesses (economic)

The gas savings reported in Table 4 translate into the total savings of 292,951 liters of gas for New York City and the consequent [33] carbon dioxide reduction of 687.26 metric tons in 2015.

The additional health benefit through exercise reported in Table 4 is translated into the total of 1.15 billions calories burnt in 2015.

In addition to the benefits above, another recent publication of ours [34] reports an indirect impact of Citi Bike deployments on the nearby commercial activity, in particular on incentivising activity in food-related businesses.

The cost-benefit estimate for the end users of Citi Bike (riders) could be done based on the sum of their economic and social benefits minus subscription costs, as described in Table 6.

| Year | Annual subscribers | Annual costs per subscriber | Annual benefits per subscriber | Benefit-cost ratio, BCR |
|---|---|---|---|---|
| **2013** | **96,125** | **$95/year** | Economic: $341.74<br>Exercise: 7,905 cal<br>Time savings: 5.78 h<br>Gas savings: 2.15 liters | 3.60 |
| **2015** | **92,781** | **$149/year** | Economic: $496.21<br>Exercise: 12,437 cal<br>Time savings: 8.19 h<br>Gas savings: 3.16 liters | 3.33 |

**Table 6.** Cost-benefit analysis for the end user (only monetary benefits accounted for).

The resulting benefit-cost ratio for the end user varies from 3.6 in 2013 to 3.33 in 2015, meaning that a Citi Bike subscription is a worthy investment for an average customer and a subscription price increase in 2015 is pretty balanced with the growing benefits. The actual BCR is even higher for those customers getting a discounted membership. A standard 10% discount is offered for Citibank customers and multiple other discounted options available for the members of the selected credit unions, New York City Housing Authority residents and others - the subscription cost could go as low as $60/year. For those users, the BCR can be as high as 8.27. The estimation of these benefit-cost ratios is done for two key Citi

Bike deployment years - in 2013 and in 2015, therefore, not taking into account present values over a longer appraisal period. The goal of benefit-cost ratio estimation in this paper is to show a sample of potential benefits for the end user, and an extended cost-benefit analysis for other stakeholders could be done separately for a longer appraisal time period.

Acknowledging the limitation of the study, it is worth mentioning that the baseline model of transportation choices in the absence of Citi Bike used for estimating marginal benefits depends on a number of assumptions on people's behavior specified above and while quantifiable uncertainty was incorporated into the estimates, the qualitative assumptions of the basic model being used are critical for the validity of the estimates. And while incorporating further detail into the transportation choice model could definitely benefit the study, at the moment even a rough estimate for the balanced multi-modal baseline scenario provides a significant step forward in understanding marginal benefits of the bikeshare deployment as compared to the single-scenario estimates available so far.

**Conclusions**

The impact of Citi Bike deployment is assessed for the two key deployment phases: July 2013 in Manhattan and Brooklyn and September 2015 in Queens, Jersey City, and additional areas of Manhattan and Brooklyn. We construct the baseline transportation model to describe transportation modes which would have been likely used to facilitate the given amount of Citi Bike ridership if Citi Bike was not available. Citi Bike trips which are unlikely to represent a rational commute choice as compared with the available alternatives are considered presumably recreational. Once the probabilistic baseline scenario is established, the direct impact of the given amount of Citi Bike rides is assessed - each Citi Bike trip provides a number of benefits in comparison with the suggested baseline alternative through the transportation costs reduction, travel time savings, additional exercise for the users and/or gas savings - multiple impacts are assessed across economic, social and environmental domains. Presumably recreational trips are excluded from the economic and environmental assessments and are only accounted for as providing additional exercise compared to a walk taken instead for the same amount of time.

The summary of all the identified impacts converted into the monetary terms allows us to conclude that the overall assessment of the Citi Bike as an urban innovation is positive. It turns out to be particularly beneficial for the end users, with an overall benefit-cost ratio from 3.33 to 8.27 depending on the subscription rate discount. The latter is often applied to the low-income population, for example residents of the subsidized housing from the NYC Housing Authority.

The study could be useful for a holistic understanding of the efficiency of the Citi Bike deployment for different types of economic, social and environmental stakeholders, including city agencies, as well as for informing future decisions on bikeshare deployments in other areas.

**Acknowledgements**
This research was funded by the Future Cities Catapult under the Performance in Use project. We thank Shefali Enaker, Adam Rae, Claudia Andrade and others from the Future Cities Catapult as well as Achilles Edwin Alfred Saxby, Richard A. Vecsler, Vishwajeet Yashwantrao Shelar, Maisha Lopa and other collaborators from NYU for their valuable suggestions and stimulating discussions.

**References**

[1] Citi Bike Monthly Operating Report (2017). Retrieved from: https://www.CitiBikenyc.com/system-data/operating-reports

[2] Fishman, E. (2015). Bikeshare: A Review of Recent Literature. Transport Reviews. Volume 36, 2016 - Issue 1: Cycling As Transport.

[3] NYC DOT Mobility Report (2016). Retrieved from: http://www.nyc.gov/html/dot/downloads/pdf/mobility-report-2016-print.pdf

[4] El- Geneidy, A. van Lierop, D., & Wasfi, R. (2015). Do people value bicycle sharing? A multilevel longitudinal analysis capturing the impact of bicycle sharing on residential sales in Montreal, Canada. Transport Policy.

[5] Wang, Lindsey, Schoner, & Harrison (2012). Modeling Bike Share Station Activity: Effects of Nearby Businesses and Jobs on Trips to and from Stations. Journal of Urban Planning and Development. 142 (1),

[6] Kisner, C. (2011). Integrating Bike Share Programs into a Sustainable Transportation System. A product of the National League of Cities, in conjunction with its Sustainability Partner, The Home Depot Foundation.

[7] The Health Risks and Benefits of Cycling in Urban Environments Compared with Car Use: Health Impact Assessment Study (2011). British Medical Journal.

[8] Sutton, M. (2009). Health and fitness converting more cyclists. Bike Biz.

[9] TNO | Knowledge for Business (2009). Retrieved from: http://www.vcl.li/bilder/518.pdf.

[10] https://www.ncbi.nlm.nih.gov/pubmed/10847255.

[11] Viechnicki, P., Khuperkar, A., Fishman, T. D., Eggers, W. D. (2015). Smart mobility: Reducing congestion and fostering faster, greener, and cheaper transportation options. Deloitte Smart Mobility Research Report.

[12] Hamilton, T.L., Wichman, C. J. (2018). Bicycle infrastructure and traffic congestion: Evidence from DC's Capital Bikeshare, Journal of Environmental Economics and Management, Volume 87, 2018, Pages 72-93, ISSN 0095-0696, https://doi.org/10.1016/j.jeem.2017.03.007.

[13] Singhvi, D.; Singhvi, S.; Frazier, P. I.; Henderson, S. G.; Mahony, E. O.; Shmoys, D. B.; and Woodard, D. B. (2015). Predicting bike usage for New York City's bike sharing system. In *Association for the Advancement of Artificial Intelligence Proceedings*.

[14] O'Mahony, Eoin and David B. Shmoys (2015). Data Analysis and Optimization for (Citi)Bike Sharing. *AAAI*.

[15] Wang, W. (2016). Forecasting bike rental demand using New York Citi Bike data. A thesis submitted in fulfilment of the requirements for the degree of MSc. In Computing (Data Analytics) in the Dublin Institute of Technology, School of Computing College of Science of Health.

[16] Stead, T. (2015). Sharing is Caring: An Analysis of Citi Bike's Discounted Program for NYCHA Residents. Pratt Institute, New York, NY.

[17] McFadden, D. (1973). Conditional logit analysis of qualitative choice behavior. In Zaremmbka, P. (ed.), Frontiers in Econometrics, Academic Press, New York.

[18] Train, K. (2003). Discrete Choice Methods with Simulation. Cambridge University Press, United Kingdom.

[19] Ben-Akiva, M. and Lerman, S. R. (1985). Discrete Choice Analysis. MIT Press, Cambridge, Massachusetts.

[20] http://web.mta.info/mta/news/hearings/2017FareTolls_proposed/proposal-eng.htm
[21] American Automobile Association (2014). Owning and operating your vehicle just got a little cheaper according to AAA's 2014 'Your Driving Costs' study.Retrieved from: http://newsroom.aaa.com/tag/driving-cost-per-mile/
[22] https://en.wikipedia.org/wiki/Sørensen–Dice_coefficient
[23] https://en.wikipedia.org/wiki/Cosine_similarity
[24] Salon, D. (2009). Neighborhoods, cars, and commuting in New York City: A discrete choice approach. *Transportation Research Part A: Policy and Practice*, *43*(2), pp.180-196.
[25] Katz, M. L., and Shapiro, C. (1985). Network Externalities, Competition, and Compatibility. *The American Economic Review*, vol. 75, no. 3, 1985, pp. 424–440. *JSTOR*, JSTOR, www.jstor.org/stable/1814809.
[26] Clements, M. T. (2004). Direct and indirect network effects: are they equivalent?, International Journal of Industrial Organization, Volume 22, Issue 5, 2004, Pages 633-645, ISSN 0167-7187, https://doi.org/10.1016/j.ijindorg.2004.01.003.
[27] http://www.bicycling.com/training/calorie-counter
[28] http://calorielab.com/burned/?mo=se&gr=17&ti=walking&q=&wt=150&un=lb&kg=68
[29] https://phys.org/news/2016-11-average-fuel-economy-high-mpg.html
[30] https://www.epa.gov/energy/greenhouse-gases-equivalencies-calculator-calculations-and-references gives an estimate of 2.346 metric tons of carbon dioxide from 1000 liters of gas.
[31] Based on EPA, AP-42: Compilation of Air Emission Factors, Table 3.3-1, Gasoline And Diesel Industrial Engines. Retrieved from: https://www3.epa.gov/ttnchie1/ap42/ch03/final/c03s03.pdf.
[32] Sustainable Streets: Strategic Plan for the New York City Department of Transportation and Beyond (2008). New York City Department of City Planning. Retrieved February 16, 2017. http://www.nyc.gov/html/dot/downloads/pdf/stratplan_compplan.pdf
[33] U.S. Environmental Protection Agency (EPA). Greenhouse Gases Equivalencies Calculator Calculations and References. Retrieved from: https://www.epa.gov/energy/greenhouse-gases-equivalencies-calculator-calculations-and-referen es
[34] Sobolevsky, S., Levitskaya, E., Chan, H., Enaker, S., Bailey, J., Postle, M., Loukachev, Y., Rolfs, M. & Kontokosta, C. (2017). Impact Of Urban Technology Deployments On Local Commercial Activity. *arXiv preprint arXiv:1712.00659*.

## Supplementary Information

### S1. The deployment

The first pilot phase in the summer of 2013 included station deployments in southern Manhattan and northern Brooklyn with the goal to eventually cover all five boroughs of New York City. Later deployment phases (summer 2015 and 2016) covered further areas in Manhattan and Brooklyn, together with some adjacent areas of Queens as well as Jersey City, New Jersey. The maps of deployment phases could be found in Figure S1. Places for deployment are based upon the feedback from the local communities and population density within a city, as well as commuting routes.

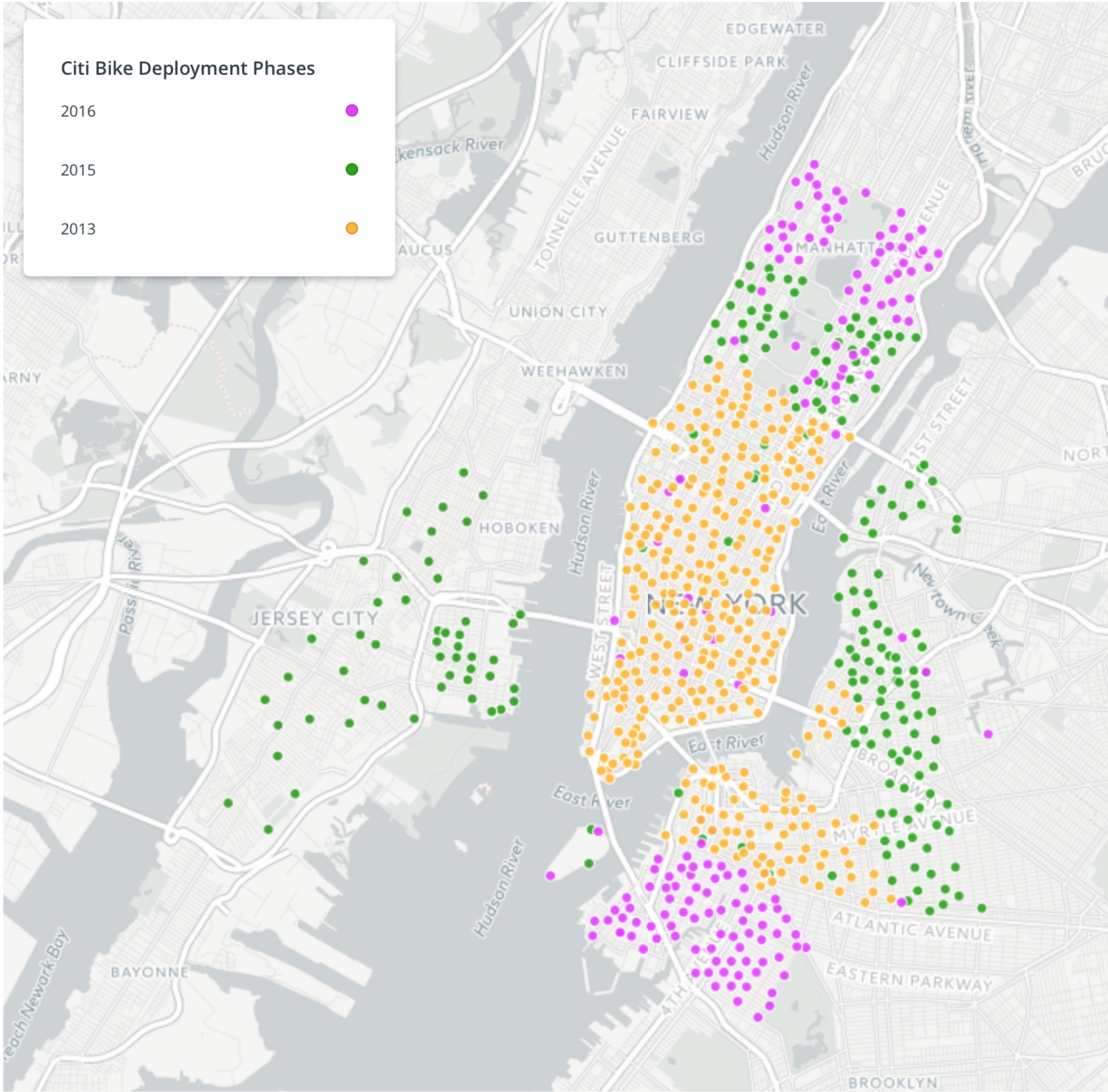

**Figure S1.** Map of Citi Bike deployment phases in New York City and Jersey City.

A bike share scheme was first proposed by the New York City Department of Transportation in 2008 in order to reduce emissions, road wear, reduce traffic congestion, and improve public health. Due to data availability, we are looking at impacts on emissions, traffic congestion and public health, as there is not enough accessible data for assessing road wear. The original operator of Citi Bike was Alta Bicycle Share, which was then succeeded by the Motivate company. Motivate currently manages all of the largest bike share systems in the United States and many of the largest systems in the world, including Ford GoBike (California Bay Area), Citi Bike (New York and Jersey City), Divvy (Chicago), CoGo Bike Share (Columbus, Ohio), Capital Bike Share (Washington, D.C., Arlington and Alexandria, Va., and Montgomery County, Md.), Hubway (Boston, Somerville, Cambridge and Brookline, Mass.), Bike Chattanooga (Tenn.) and BIKETOWN (Portland, OR)[1].

The system in New York City was supposed to start in 2011 but was postponed to 2012 because of uncertainties related to deployment locations of stations. The city wanted to locate Citi Bike stations on sidewalks and public plazas, but there were places where stations would take up parking space[2]. The 2012 start for the first deployment was later delayed to 2013 because of software problems[3]. Citi Bike was finally launched on May 27, 2013.

## S2. The data sources

### NYC Bike Routes[4]

Dataset is created by NYC Department of Transportation, Bicycle and Greenway Program with contributions from NYC Department of City Planning, NYC Department of Parks and Recreation. Bicycle Routes is a line representation of the NYC bicycle network which contains information regarding the bicycle facilities as well as the relevant street information. The purpose of this project is to identify existing bicycle lanes/paths and to provide to other agencies and the public information about the city's bicycle network in a GIS format. This dataset is updated annually. This is a vector data of geometry type: polyline. Spatial Reference: Geographic Coordinate Reference: GCS_North_American_1983 Projection: NAD_1983_StatePlane_New_York_Long_Island_FIPS_3104_Feet

**Data structure:** street, SegmentID, boro, fromstreet, tostreet, onoffst, allclasses, Instdate, moddate, comments, bikedir, lanecount, ft_facility, Tf_facility.

[1] https://www.Citi Bikenyc.com/about
[2] Haughney, C. (2011). "New York Chooses Company to Run Bike-Share Program". The New York Times.
[3] Bernstein, Andrea (2012). "Bloomberg: NYC Bike Share Delayed Until Spring (UPDATED)". WNYC.
[4] Available through NYC Open Data Portal: https://data.cityofnewyork.us/Transportation/Bike-Routes/7vsa-caz7/data

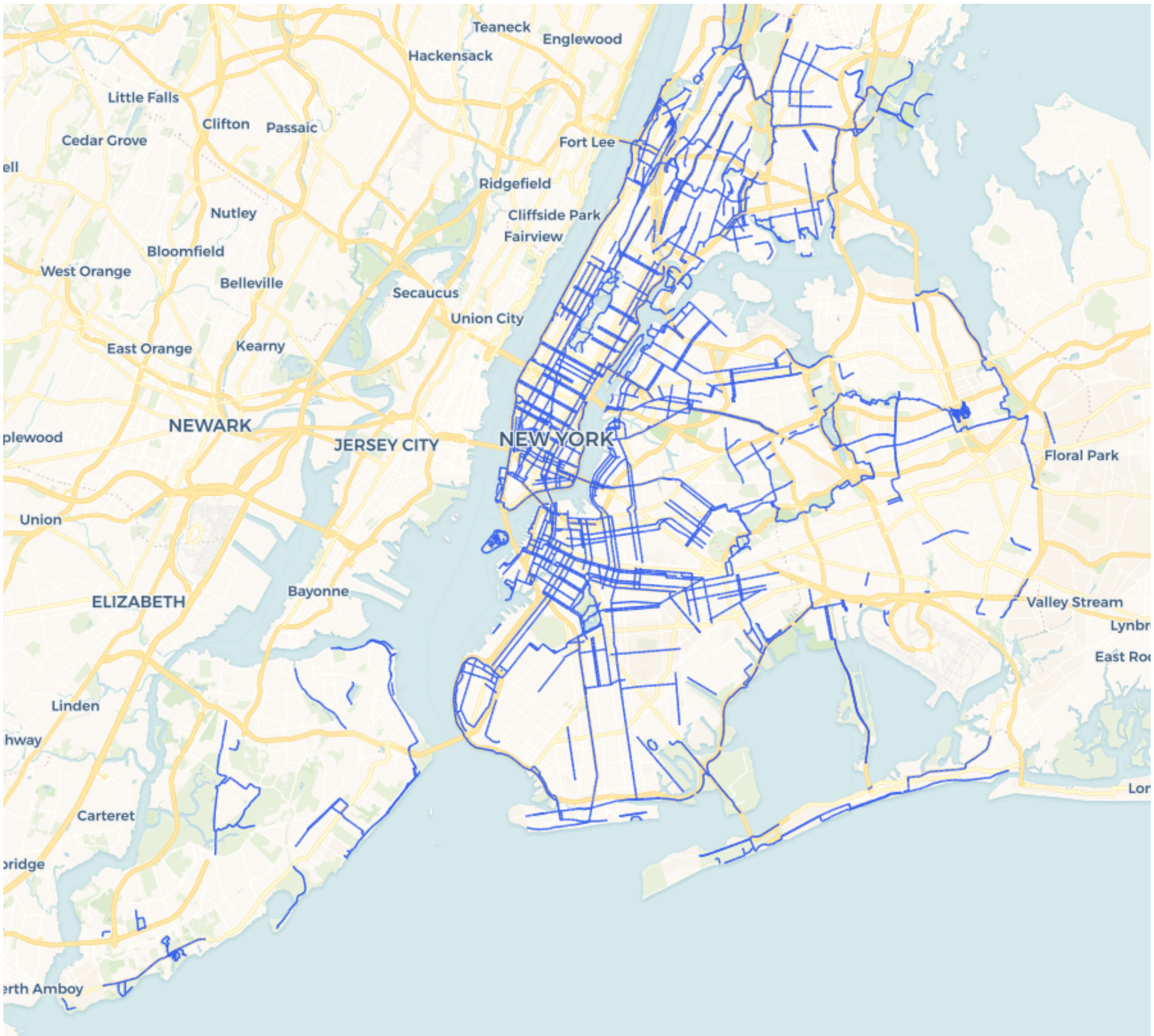

**Figure S2.** Locations of bike lanes and routes throughout the New York City.

## Taxi Usage & Locations[5]

The yellow and green taxi trip records include fields capturing pick-up and drop-off dates/times, pick-up and drop-off locations, trip distances, itemized fares, rate types, payment types, and driver-reported passenger counts. The data used in the attached datasets were collected and provided to the NYC Taxi and Limousine Commission (TLC) by technology providers authorized under the Taxicab & Livery Passenger Enhancement Programs (TPEP/LPEP). The trip data was not created by the TLC, and TLC makes no representations as to the accuracy of these data.

The For-Hire Vehicle ("FHV") trip records include fields capturing the dispatching base license number and the pick-up date, time, and taxi zone location ID. These records are generated from the FHV Trip

[5] Available on the website of NYC Taxi and Limousine Commission: http://www.nyc.gov/html/tlc/html/about/trip_record_data.shtml

Record submissions made by bases. Note: The TLC publishes base trip record data as submitted by the bases, and cannot guarantee or confirm their accuracy or completeness. Therefore, this may not represent the total amount of trips dispatched by all TLC-licensed bases. The TLC performs routine reviews of the records and takes enforcement actions when necessary to ensure, to the extent possible, complete and accurate information.

**Data structure - Yellow taxi:** VendorID, tpep_pickup_datetime, tpep_dropoff_datetime, Passenger_count, Trip_distance, Pickup_longitude, Pickup_latitude**,** RateCodeID, Store_and_fwd_flag, Dropoff_longitude, Dropoff_ latitude, Payment_type, Fare_amount, Extra, MTA_tax, Improvement_surcharge, Tip_amount, Tolls_amount, Total_amount

**Data structure - Green taxi:** VendorID, lpep_pickup_datetime, lpep_dropoff_datetime, Passenger_count, Trip_distance, Pickup_longitude, Pickup_latitude, RateCodeID, Store_and_fwd_flag**,** Dropoff_longitude, Dropoff_ latitude, Fare_amount, Extra, MTA_tax, Improvement_surcharge**,** Tip_amount, Tolls_amount**,** Total_amount, Trip_type

**Data structure - FHV:** Dispatching_base_num, Pickup_date, locationID

**Uber Usage & Locations[6]**

This data contains over 4.5 million Uber pickups in New York City from April to September 2014, and 14.3 million more Uber pickups from January to June 2015. Trip-level data on other for-hire vehicle (FHV) companies is also included. All the files are as they were received on August 3, Sept. 15 and Sept. 22, 2015.

FiveThirtyEight obtained the data from the NYC Taxi & Limousine Commission (TLC) by submitting a Freedom of Information Law request on July 20, 2015. The data has the following columns: Dispatching_base_num, Pickup_date, Affiliated_base_num, locationID. The taxi Zone and Borough are associated with each locationID.

**Subway Locations[7]**

Locations of New York City subway stations.
**Data structure:** URL, OBJECTID, NAME, the_geom, LINE, NOTES.

**Subway Usage[8]**
Turnstile data on entries and exits of New York City subway stations.

---

[6] FiveThirtyEight obtained the data from the NYC Taxi & Limousine Commission (TLC) by submitting a Freedom of Information Law request on July 20, 2015: https://github.com/fivethirtyeight/uber-tlc-foil-response

[7] Available through NYC Open Data Portal: https://data.cityofnewyork.us/Transportation/Subway-Stations/arq3-7z49/data

[8] Available on the website of Metropolitan Transportation Authority (MTA): http://web.mta.info/developers/turnstile.html

**Data structure:** C/A, UNIT, SCP, STATION, LINENAME, DIVISION, DATE, TIME, DESc, ENTRIES, EXIST.

## Citi Bike System Data[9]

**Data structure:** Trip Duration (seconds), Start Time and Date, Stop Time and Date, Start Station Name, End Station Name, Station ID, Station Lat/Long, Bike ID, User Type (Customer = 24-hour pass or 3-day pass user; Subscriber = Annual Member), Gender (Zero=unknown; 1=male; 2=female), Year of Birth.

This data has been processed to remove trips that are taken by staff as they service and inspect the system, trips that are taken to/from any of our "test" stations (which we were using more in June and July 2013), and any trips that were below 60 seconds in length (potentially false starts or users trying to re-dock a bike to ensure it's secure).

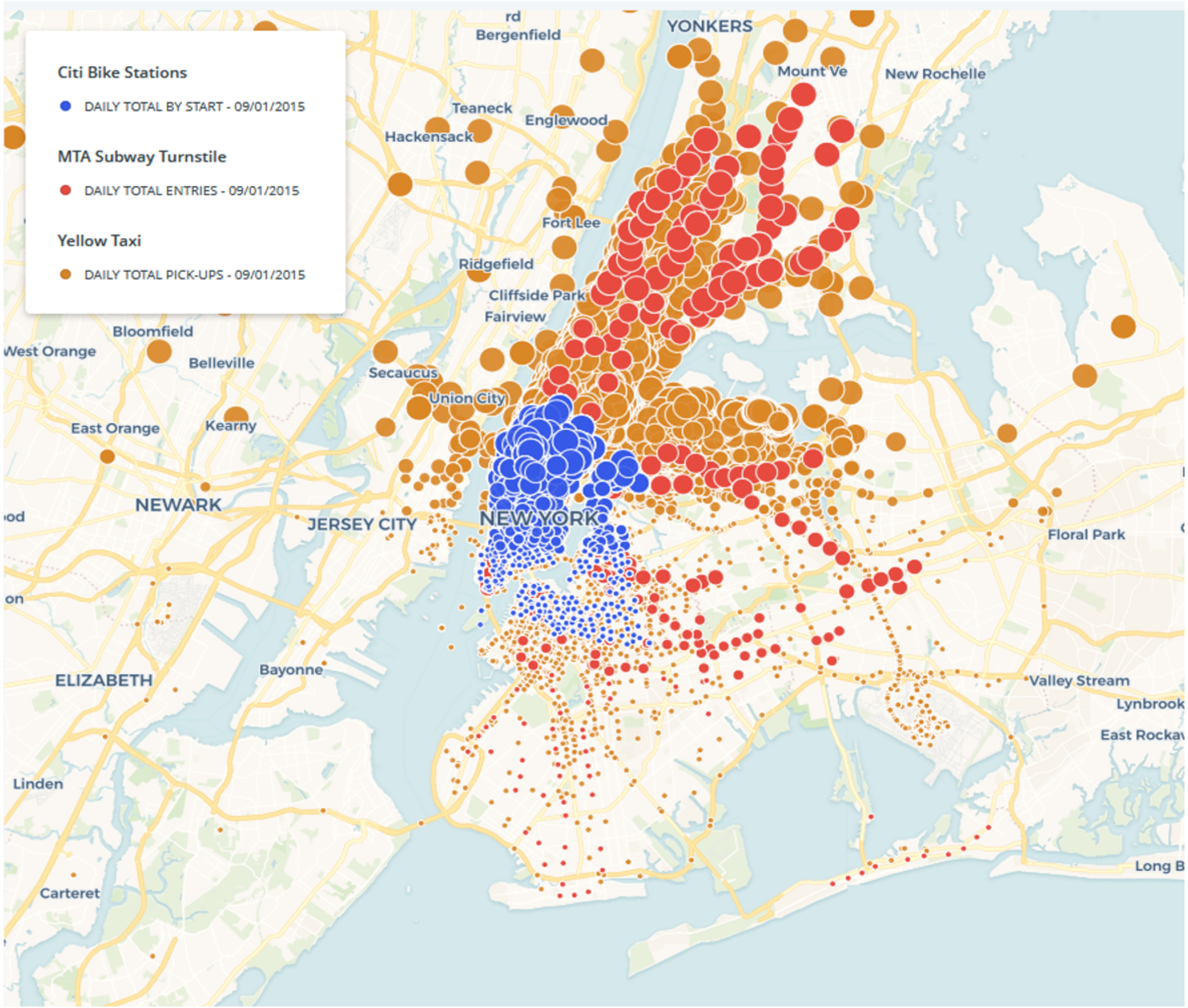

**Figure S3.** A sample of daily statistics and geographic distribution per transportation mode in New York City for September 01, 2015. Larger bubbles indicate a larger amount of use.

9 Available from: https://www.Citi Bikenyc.com/system-data

**NYC-specific data from the U.S. Census Bureau**[10]

Data products of U.S. Census Bureau:

- Censuses
    - The Decennial Census is the once-a-decade population and housing count of all 50 states, the District of Columbia, Puerto Rico and the Island Areas as required by the U.S. Constitution. The results of the decennial census determine the number of seats for each state in the U.S. House of Representatives and are used to draw congressional and state legislative districts and to distribute more than $675 billion in federal funds each year.
    - The Economic Census measures the nation's economy every five years, providing vital statistics for virtually every industry and geographic area in the country.
    - The Census of Governments provides comprehensive data about the nearly 90,000 state and local governments in the nation every five years.

- Surveys
    - The American Community Survey (ACS) is an ongoing annual survey that shows what the U.S. population looks like and how it lives. The ACS helps communities decide where to target services and resources.
    - Demographic surveys measure income, poverty, education, health insurance coverage, housing quality, crime victimization, computer usage, and many other subjects.
    - Economic surveys are conducted monthly, quarterly, and yearly. They cover selected sectors of the nation's economy and supplement the Economic Census with more-frequent information about the dynamic economy. These surveys yield more than 400 annual economic reports, including principal economic indicators.
    - Sponsored surveys are demographic and economic surveys that we conduct for other government agencies. They include the Current Population Survey, the National Health Interview Survey, and the National Survey of College Graduates.

- Population Estimates and Projections
    - Every year, the Census Bureau publishes population estimates and demographic components of change, such as births, deaths, and migration. This data can be sorted by characteristics such as age, sex, and race, as well as by national, state, and county location.
    - The Census Bureau estimates population size and characteristics. Population projections are based on future demographic trends, including births, life expectancy, and migration patterns.

**Longitudinal Employment Household Dynamics (LEHD)**[11]

LEHD makes available several data products that may be used to research and characterize workforce dynamics for specific groups. These data products include online applications, public-use data, and restricted-use microdata. The Quarterly Workforce Indicators (QWI) and LEHD Origin-Destination

[10] Available from: https://www.census.gov/
[11] Available from: https://lehd.ces.census.gov/data/

Employment Statistics (LODES) data are available online for public use. Confidential microdata are available to qualified researchers with approved projects through restricted access use in Census Research Data Centers.

LEHD Origin-Destination Employment Statistics (LODES) used by OnTheMap are available for download below. Data files are state-based and organized into three types: Origin-Destination (OD), Residence Area Characteristics (RAC), and Workplace Area Characteristics (WAC), all at census block geographic detail. Data is available for most states for the years 2002–2015.

**American Community Survey**[12]

American Community Survey is an ongoing annual survey that shows what the U.S. population looks like and how it lives. The ACS helps communities decide where to target services and resources.

Key Data Products:

- Data Profiles - Provide broad social, economic, housing, and demographic profiles.
- Comparison Profiles - Similar to Data Profiles but show data side-by-side from the five most recent years of the ACS.
- Selected Population Profiles - Provide broad social, economic, housing, and demographic profiles for a large number of race, ethnic, ancestry, and country/region of birth groups.
- Ranking Tables - Provide state rankings of estimates across 86 key variables.
- Subject Tables - Similar to Data Profiles but include more detailed ACS data, classified by subject.
- Detailed Tables - Provide access to the most detailed ACS data and cross tabulations of ACS variables.
- Geographic Comparison Tables - Compare geographic areas other than states (e.g., counties or congressional districts) for key variables.
- Summary Files - Provide access to the Detailed Tables through a series of comma-delimited text files.
- Public Use Microdata Sample (PUMS) Files - Provide access to ACS microdata for data users with statistical software experience.

Population Category: Age, Ancestry, Citizenship Status, Commuting (Journey to Work) and Place of Work, Disability Status, Educational Attainment and School Enrollment, Employment Status, Fertility, Grandparents as Caregivers, Health Insurance Coverage, Hispanic or Latino Origin, Income and Earnings Industry, Occupation, and Class of Worker, Language Spoken at Home, Marital History, Marital Status, Migration/Residence 1 Year Ago, Period of Military Service, Place of Birth, Poverty Status, Race, Relationship to Householder, Sex, Undergraduate Field of Degree, VA Service-Connected Disability Status, Veteran Status, Work Status Last Year, Year of Entry

[12] Available from: https://www.census.gov/programs-surveys/acs/

<u>Housing Category</u>: Acreage and Agricultural Sales, Bedrooms, Computer and Internet Use, Food Stamps/Supplemental Nutrition Assistance Program (SNAP), House Heating Fuel, Kitchen Facilities, Occupancy/Vacancy Status, Occupants Per Room, Plumbing Facilities, Rent, Rooms, Selected Monthly Owner Costs, Telephone Service Available, Tenure (Owner/Renter), Units in Structure, Value of Home, Vehicles Available, Year Householder Moved Into Unit, Year Structure Built.

**S3. Routing statistics**

The projected travel time and distance for each transportation mode according to each origin-destination pair of each Citi Bike trip recorded is estimated based on a random sample of trajectories scraped from Google Maps API[13] across each of the considered areas for several typical times of the day/week (9am, 2pm, 6pm weekdays and 2pm weekends, including most common and pessimistic scenarios for driving, 100 trajectories per each time/area). The average travel time and distance traveled per mile of direct distance is obtained by running a linear regression for the actual observed times and distances versus the direct Euclidean distance between the origin and destination. This way the travel time and distance estimates for each origin-destination pair and transportation mode also incorporate uncertainty from this regression. The results for different boroughs of NYC and for Jersey City, different transportation modes and times of the day/week are reported in the tables S1-S3 below.

| | Walk, min/mile | Bike, min/mile | Public Transit, min/mile | Drive, min/mile | Drive, pessimistic, min/mile | Drive distance, mile/mile |
|---|---|---|---|---|---|---|
| Weekday, 9am | 23.477±0.145 | 6.917±0.079 | 10.453±0.335 | 7.109±0.126 | 10.686±0.182 | 1.267±0.015 |
| Weekday, 2pm | | | 10.753±0.345 | 6.242±0.125 | 8.631±0.169 | 1.354±0.022 |
| Weekday, 6pm | | | 10.428±0.343 | 6.491±0.124 | 9.376±0.179 | 1.307±0.017 |
| Weekend, 2pm | | | 11.441±0.380 | 6.235±0.124 | | 1.344±0.022 |

Table S1. Commute time and distance per mile of geographical distance in Jersey City

[13] https://developers.google.com/maps/

| | Walk, min/mile | Bike, min/mile | Public Transit, min/mile | Drive, min/mile | Drive, pessimistic, min/mile | Drive distance, mile/mile |
|---|---|---|---|---|---|---|
| Weekday, 9am | 25.658±0.337 | 9.845±0.144 | 17.139±0.467 | 10.136±0.246 | 15.776±0.384 | 1.790±0.049 |
| Weekday, 2pm | | | 16.962±0.480 | 9.622±0.226 | 14.525±0.337 | 1.759±0.045 |
| Weekday, 6pm | | | 16.922±0.429 | 10.603±0.263 | 16.656±0.407 | 1.728±0.036 |
| Weekend, 2pm | | | 16.959±0.471 | 9.040±0.226 | | 1.793±0.049 |

Table S2. Commute time and distance per mile of geographical distance in Brooklyn

| | Walk, min/mile | Bike, min/mile | Public Transit, min/mile | Drive, min/mile | Drive, pessimistic, min/mile | Drive distance, mile/mile |
|---|---|---|---|---|---|---|
| Weekday, 9am | 23.670±0.182 | 8.942±0.103 | 10.743±0.404 | 9.661±0.201 | 16.062±0.356 | 1.449±0.031 |
| Weekday, 2pm | | | 10.785±0.410 | 9.365±0.212 | 15.414±0.383 | 1.529±0.036 |
| Weekday, 6pm | | | 10.872±0.395 | 9.211±0.208 | 15.344±0.366 | 1.461±0.034 |
| Weekend, 2pm | | | 10.684±0.414 | 9.441±0.223 | | 1.500±0.037 |

Table S3. Commute time and distance per mile of geographical distance on Manhattan